\documentclass[12pt,english,aps,preprintnumbers,amsmath,amssymb]{revtex4}
\usepackage{graphicx}

\makeatletter
\@ifundefined{textcolor}{}
{%
 \definecolor{BLACK}{gray}{0}
 \definecolor{WHITE}{gray}{1}
 \definecolor{RED}{rgb}{1,0,0}
 \definecolor{GREEN}{rgb}{0,1,0}
 \definecolor{BLUE}{rgb}{0,0,1}
 \definecolor{CYAN}{cmyk}{1,0,0,0}
 \definecolor{MAGENTA}{cmyk}{0,1,0,0}
 \definecolor{YELLOW}{cmyk}{0,0,1,0}
 }

\def\Eu8{Eu(Fe$_{0.81}$Co$_{0.19}$)$_2$As$_2$}
\def\EuFeCoAs{Eu(Fe$_{1-x}$Co$_{x}$)$_2$As$_2$}
\def\CaCo_s2{Ca(Fe$_{0.96}$Co$_{0.04}$)$_2$As$_2$}
\makeatother

\usepackage{babel}
\begin{document}
Submitted to Nature Materials, Nov 30, 2011

\title{Tuning superconductivity by magnetic fields in \Eu8}

\author{Vinh Hung Tran, Tomasz A. Zaleski, Zbigniew Bukowski, Lan Maria Tran,
Andrzej J. Zaleski}

\affiliation{Institute of Low Temperature and Structure Research, Polish Academy
of Sciences, P.O. Box 1410, 50-422 Wroc{\l{}}aw, Poland}

\maketitle
\textbf{The distinct difference between BCS-type and unconventional
triplet superconductivity manifests itself in their response to external
magnetic fields. An applied field easily extinguishes s-wave singlet
superconductivity by both the paramagnetic or orbital pair-breaking
effects. However, it hardly destroys triplet state because the paramagnetic
effect, owing to spins of the Cooper pairs readily aligned with the
field, is not so efficacious. This suggests that the triplet superconductivity
may be affected mostly by the orbital effect. Conversely, if one can
break down the orbital effect then one can recover the superconductivity.
Here, we show that superconductivity can be induced with magnetic
fields applied parallel to the \emph{ab} plane of crystals of the magnetic
\Eu8~superconductor. We argue that the tuning superconductivy may
be actuated by relative enhancement of ferromagnetic interactions
between the Eu$^{2+}$ moments lying in adjacent layers  and removal
of their canting toward \emph{c} axis that is present in zero field.}

\begin{figure}[h]
 \includegraphics[scale=0.6]{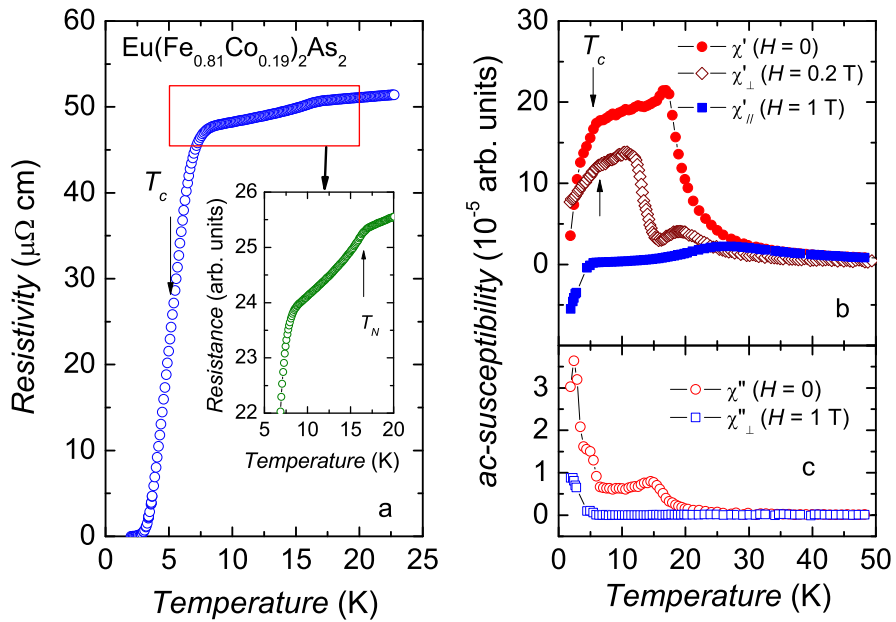} \caption{a) Temperature dependence of the ac-electrical resistivity of Eu(Fe$_{0.81}$Co$_{0.19}$)$_{2}$As$_{2}$
measured with a current of 5 mA. The superconducting transition temperature
$T_{c}\sim$ 5.1 K is defined as a midpoint of the resistivity jump.
The inset shows a change in the resistivity slope around $T_{N}$
in an enlarged scale. b) The real component $\chi^{\prime}(T)$ and
c) the imaginary component $\chi^{\prime\prime}(T)$ of ac-magnetic
susceptibility as a function of temperature.}

\label{Fig_Resi_Chi}
\end{figure}

\Eu8~ is one of the members of solid solutions \EuFeCoAs~ crystallizing
in the tetragonal ThCr$_{2}$Si$_{2}$-type structure (space group
I4/mmm) at room temperature. In the unit cell of the parent EuFe$_{2}$As$_{2}$
compound, the Fe$^{2+}$ ions distributed on the Z = 0.25 layers exhibit
a spin-density wave (SDW) ordering below 190 K, while the Eu$^{2+}$
ions located on the Z = 0 layers order antiferromagnetically below
19 K. The magnetic moments of Fe$^{2+}$ and Eu$^{2+}$ are parallel
to one another, and are confined to the \emph{ab} plane~\cite{Marchand_78,Ren_08,Jeevan_08,Wu_09}.
The suppression of the SDW state in EuFe$_{2}$As$_{2}$, by hydrostatic
pressure~\cite{Miclea_09,Terashima_09} or by suitable chemical substitution,
i.e., \EuFeCoAs~ with 0.18 $<x<$ 0.3~\cite{Zheng_09,Matusiak_11}, results
in the emergence of the superconductivity. It was also shown that
the occurrence of superconductivity does not significantly change
the magnetic ordering temperature $T_{N}$ of the Eu$^{2+}$ magnetic
ions.\\
\par ac electrical resistivity ($\rho(T)$) and ac magnetic susceptibility
($\chi^{\prime}(T)$, $\chi^{\prime\prime}(T)$) of the \Eu8~single crystal, \cite{Tran_2011} presented in Fig. \ref{Fig_Resi_Chi}
exemplify the coexistence of magnetism and superconductivity. In the
studied compound, the magnetic ordering of the Eu$^{2+}$ ions sets
in at the N\'{e}el temperature $T_{N}\sim$ 16.5 K and superconductivity
at the critical temperature $T_{c}\sim$ 5.1 K.

Note that as external field is applied parallel to the \emph{ab} plane the susceptibility reveals two well separated maxima. The low temperature and high temperature anomalies correspond to antiferromagnetic and ferromagnetic components, respectively. For fields above 1 T, $\chi_{\parallel}^{\prime}(T)$ and $\chi_{\perp}^{\prime}(T)$ behave similarly, e.g., showing a negative value below $T_c$ and a maximum at $\sim$ 25 K conforming the field-induced ferromagnetic arrangement of Eu$^{2+}$. We pay attention to the presence of a dissipative
process in $\chi^{\prime\prime}(T)$ below $T_{N}$
(Fig. \ref{Fig_Resi_Chi} c), which is in agreement with the M\"{o}ssbauer data~
\cite{Blachowski_11}, where the magnetic moments of Eu$^{2+}$
have been  designated to be canted from the \emph{c} axis by an angle of 60$^{\circ}$.
Because of the magnetic interactions, the Eu$^{2+}$ layers are still
expected to be weak conducting layers, and therefore the \emph{c} axis
remains the worse conducting direction than the \emph{ab} plane.
\begin{figure}[h]
 \includegraphics[scale=0.6]{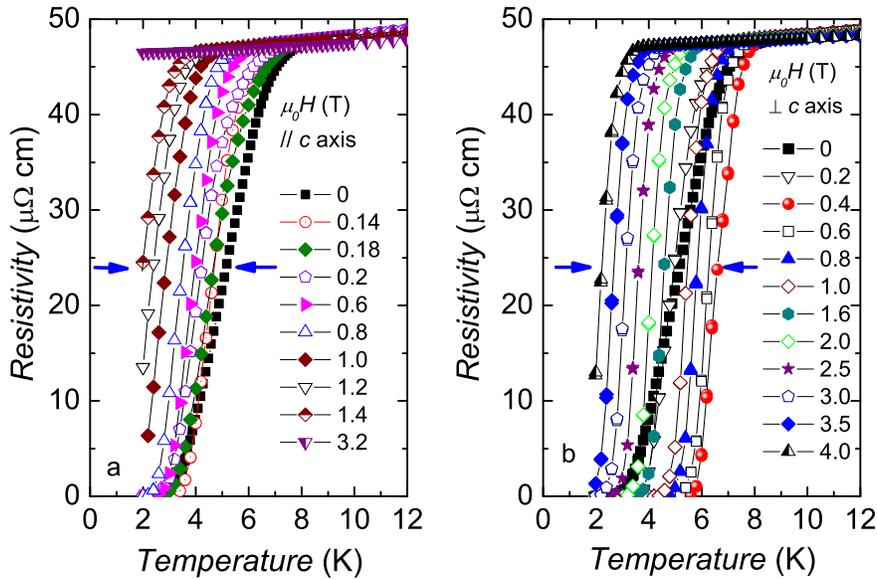} \caption{The resistivity as a function of temperature for a) $H\|c$ and b)
$H_{\perp}c$. The arrows indicate midpoints of the resistivity curves.
A perpendicular field of 0.4 T induces superconductivy by shifting $T_{c}^{R=0}$
= 2.2 K (at 0 T) up to 5.6 K.}

\label{Fig_Resi_TempH}
\end{figure}

The field-dependent resistivity around the superconducting transition
is shown in Fig. \ref{Fig_Resi_TempH}. For $H\|c$, $T_{c}$ and
the superconducting transition width, defined as $\Delta T_{c}=T90\%-T10\%$,
where $T90\%$ and $T10\%$ are the temperatures corresponding to
90$\%$ and 10$\%$ of the resistivity jump, decrease with increasing field.
However, $\Delta T_{c}$ persists with a large value of 1.7 K at fields above 1
T. It is well known that the relative domination of the two pair-breaking
effects determines the order of the phase transition in a type-II
superconductor. The dominating orbital pair-breaking is usually associated
with a second order phase transition while the dominating paramagnetic
pair-breaking is related with a first order phase transition. Owing to a regular dependence of $\Delta T_{c}(H)$ observed for \emph{$H\|c$}, no
sudden change in the pair-breaking mechanism should be expected.
\begin{figure}[h]
 \includegraphics[scale=0.6]{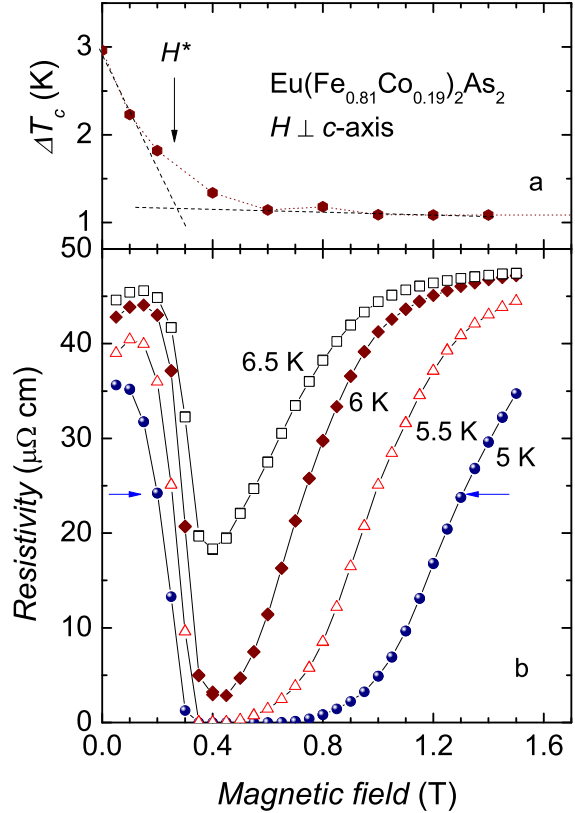} \caption{a) The superconducting transition width $\Delta T_{c}$ and b) The
electrical resistivity of \Eu8~ at 5, 5.5, 6 and 6.5 K as a function
of $H_{\perp}c$. The dashed lines in a) are linear extrapolations
of $\Delta T_{c}$ in low- and high-field regimes. The intersection
of these lines corresponds to $H^{*}$. }

\label{Fig_Resi_DTc_H}
\end{figure}

A more salient feature of the resistivity is observed under magnetic
fields perpendicular to the \emph{c} axis (Fig. \ref{Fig_Resi_TempH} b).
Here we want to point out two experimental facts which indicate important
correlations between them. One is a field-induced superconductivity
in a wide field range. It is clear seen from the figure that the zero-resistance
point $T_{c}^{R=0}$ as well as $T_{c}$ shift towards higher temperatures
for a range of fields 0.27 - 1 T. The other fact should
be noted is that $\Delta T_{c}$ rapidly narrows with increasing magnetic
fields, i.e., $\Delta T_{c}$ amounting to 2.96 K at zero field decreases
down to 1.34 K at 0.4 T (see Fig. \ref{Fig_Resi_DTc_H} a) and levels
off to 1.07 K at fields above 0.6 T. The sharpening transition with increasing
fields might be associated with a change in the pair-breaking mechanisms,
e.g., from the orbital to dominant paramagnetic one. We believe that
the suppression of the orbital pair-breaking effect may explain the
field-induced superconductivity in \Eu8. Possible other mechanisms
accounting for field-induced superconductivity will be discussed below.
The field-induced superconductivity is more evident in
the isothermal resistivity measurements shown in Fig. \ref{Fig_Resi_DTc_H}
b. The normal state of the studied sample in zero field for temperatures
between 5 K and 6.5 K can be pushed into a superconducting state via
applying a suitable magnetic strength of about $H^{*}\sim$ 0.27 T,
i.e., the field strength presumably necessary to abate the orbital
pair-breaking effect.
\begin{figure}[h]
 \includegraphics[scale=0.6]{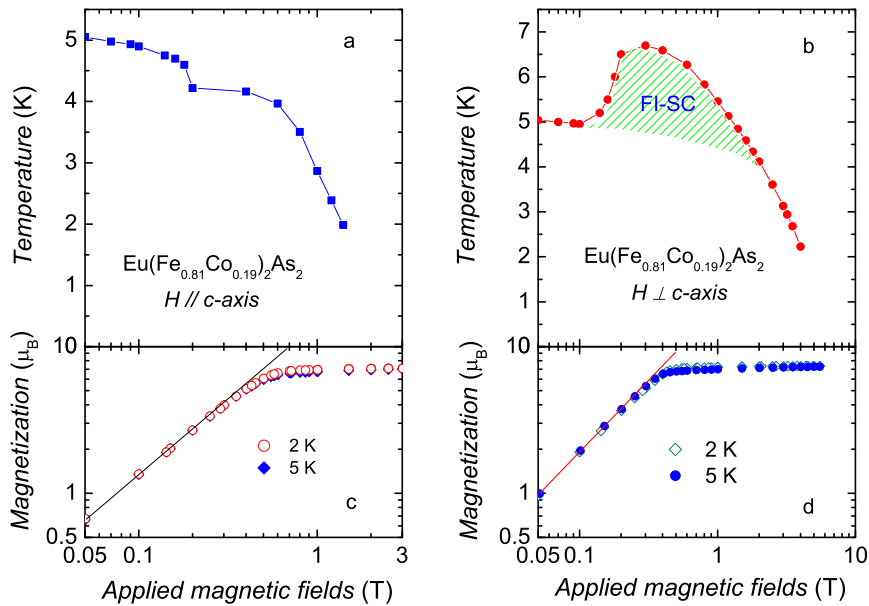} \caption{a) and b) The field-temperature phase diagram for $H\|c$ and $H\perp c$,
respectively. FI-SC denotes the area of the field-induced superconductivity.
c) and d) Magnetization at 2 and 5 K versus $H\|c$ and $H\perp c$,
respectively. }

\label{Fig_PlotHTemp}
\end{figure}

The field dependencies of $T_{c}$ are summarized in temperature versus
magnetic field phase diagrams shown in Fig. \ref{Fig_PlotHTemp} a
and Fig. \ref{Fig_PlotHTemp} b. In the vicinity of $T_{c}$, one
find a linearity between \emph{H} and $T_{c}$ , and thus one estimates
the initial slope of the $dH/dT|_{T=T_{c}}$ = -0.33 T/K and -0.49 T for $H\|c$ and $H\perp c$, respectively. Using the
Werthamer-Helfand-Hohenberg formula for a dirty limit\cite{WHH}:
\begin{equation}
H_{c2}^{orb}=-0.693T_{c}\frac{dH_{c2}}{dT}|_{T=T_{c}}
\end{equation}
 we evaluated the orbital pair-breaking fields $H_{c2}^{orb, \| c}$ = 1.2
T and $H_{c2}^{orb, \perp c}$ = 1.8 T in absence of any paramagnetic limitation. On the other hand, Clogston has shown that the paramagnetically limited upper critical field should be given by:\cite{Clogston}
\begin{equation}
H_{po}=1.84T_{c}.\label{Clogston}
\end{equation}
 For $T_{c}$ = 5.1 K and using Eq. \ref{Clogston} we estimated $H_{po}$
to be 9.5 T. An extrapolation of $T_{c}(H)$ to \emph{T} = 0 yields
the value of the upper critical field $H_{c2}(0)\sim$ 2.7 T for $H\|c$
and 6.5 T for $H_{\perp}c$. A comparison of these upper critical
fields implies that the $T_{c}(H)$ curve in the $H\|c$ configuration
is mainly governed by the orbital pair-breaking effect but a dominating
paramagnetic effect is taken down for $H_{\perp}c$. The absence of the orbital
limit for high field strength $H_{\perp}c$ is consistent with the behaviour
of $\Delta T_{c}(H)$ considered above.

The dc-magnetization (\emph{M}) data collected at 2 and 5 K (Fig. \ref{Fig_PlotHTemp} c and d) exhibit a change in the slope at about 0.2 T, indicative of spin reorientation towards the magnetic field direction. Above 1 T,
corresponding to a ferromagnetic state, $M_{\|}$ and $M_{\perp}$
attain respective value of 7.14 and 7.37 $\mu_{B}$/f.u, slightly
larger than 7 $\mu_{B}$ expected for Eu$^{2+}$. A possible contribution from
polarized Fe$^{2+}$ moments should be checked in future studies.
The fact that the superconductivity coexisting with the ferromagnetic
ordering over a wide range of magnetic field strongly suggests its
unconventional character, presumably of a triplet state with net spin
\emph{S} = 1 parallel to \emph{ab} plane. The other evidences of unconventional
superconductivity come from the observation of the field-induced superconductivity
and a large anisotropy of $T_{c}(H)$.
\begin{figure}[h]
 \includegraphics[scale=0.6]{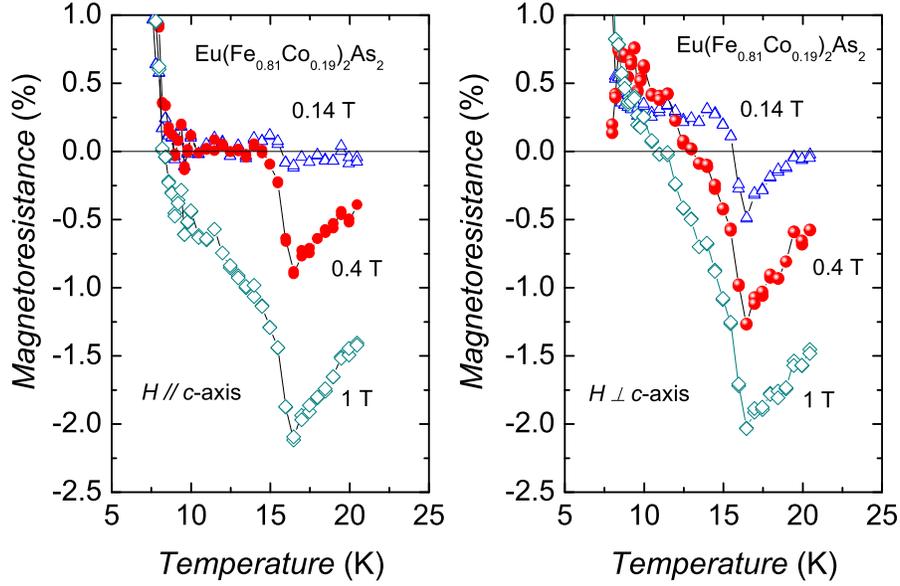} \caption{Temperature dependence of the magnetoresistance.}

\label{Fig_PlotMRTemp}
\end{figure}

The undubitable difference in the magnetoresistance (MR) for $H\|c$
and $H\perp c$ (Fig. \ref{Fig_PlotMRTemp}) demonstrates a close
relationship between the field-induced superconductivity and the disturbance
of the antiferromagnetic Eu$^{2+}$ sublattice. The spin-flip process
of the Eu$^{2+}$ moments in the adjacent layers certainly accompanies
modification of magnetic interactions with the Fe orbitals. It seems
likely that this scenario may ascribe to a factor making the orbital
pair-breaking effect insignificant.

To the best of our knowledge, there are only a few observations of field-induced
superconductivity in URhGe~\cite{Levy_05}, Eu$_{x}$Sn$_{1-x}$Mo$_{6}$S$_{8}$~
\cite{Meul_84} and $\lambda$-(BETS)$_{2}$FeCl$_{4}$~\cite{Uji_01}.
URhGe is a special case, since the field-induced superconductivity
may happen in neighborhood of a quantum transition under high magnetic
field \cite{Levy_05}. The behaviour of the second compound has been
explained by the Jaccarino-Peter effect \cite{JP}, i.e. a compensation
between the external field and the internal field created by the the polarization
of magnetic ions. For $\lambda$-(BETS)$_{2}$FeCl$_{4}$, besides possible Jaccarino-Peter compensation effect, the low dimensionality of the electronic system has been evoked for the interpretation \cite{Uji_01}.
In \Eu8, the reason for inducing the superconductivity
may be the following. In the strongly anisotropic systems, the orbital
pair-breaking effect may be suppressed if the magnetic fields is applied
parallel to the conductive direction \cite{Klemm_75}. As a result,
since movement of the electrons perpendicular to the field is limited,
the effect of the field is strongly suppressed, so is the orbital effect.
However, even small component of the field/magnetisation parallel to the $c$ direction may be sufficient to cause the orbital effect and break down the superconductivity completely. In the present case, the small component may result from canting of the Eu$^{2+}$ moments toward the c axis in zero field. Furthermore, the magnetic field applied within $ab$ planes polarizes Eu$^{2+}$ ferromagnetically along the field direction, aligning the Eu$^{2+}$ moments within the \emph{ab} planes, and therefore the superconductivity can be restored. We may add that the comparison of \Eu8~with a non-magnetic superconducting \CaCo_s2~reference, adopting the same crystal structure and similar chemical stoichiometry but with different magnetic sublattices, \cite{Tran_2011} strongly supports our interpretation that the field-induced superconductivity in \Eu8~is essentially associated with the interplay between magnetic interactions and orbital pair-breaking effect.
\par We have revealed that in the \Eu8~single crystals external magnetic field applied perpendicular to the \emph{c} axis leads to emergence of superconductivity. In order to explain the finding we have discussed several mechanisms, but the most plausible is that orbital pair-breaking effect can be abated by enhancement of the ferromagnetic interactions of the Eu$^{2+}$ moments. We believe that the field tuning superconductivity opens new possibilities towards fabrication of field-controlled devices.\\

\end{document}